\documentclass[epjST_arXiv]{svjour_arXiv}
\usepackage{amsmath}
\usepackage{graphicx,amssymb,amsfonts,latexsym,color,dcolumn,bm}
\usepackage{hyperref}

\newcommand{\rrho}{\rho_\theta^*}

\begin{document}

\title{A toy model for the dipolar-induced resonance in quasi-one-dimensional systems}

\author{
Nicola Bartolo\inst{1,2}\fnmsep\thanks{\email{nicola.bartolo@univ-paris-diderot.fr}}
\and
David J. Papoular \inst{1}
\and
Alessio Recati \inst{1}
\and
Chiara Menotti \inst{1}}
\institute{
INO-CNR BEC Center and Dip. di Fisica, Universit\`{a} di Trento, Povo, Italy
\and
Lab. C. Coulomb (L2C), UMR 5221 CNRS-Univ. Montpellier 2, Montpellier, France}

\abstract{
We discuss the properties of the effective dipolar interaction for two particles tightly confined along a one-dimensional tube, stressing the emergence of a single dipolar-induced resonance in a regime for which two classical dipoles would just repel each other.
We present a toy-model potential reproducing the main features of the effective interaction: a non-zero-range repulsive potential competing with an attractive contact term.
The existence of a single resonance is confirmed analytically.
The toy model is than generalized to investigate the interplay between dipolar and contact interaction, giving an intuitive interpretation of the resonance mechanism.}

\maketitle

\section{Introduction}\label{SecIntro}
Thanks to a high degree of experimental control, cold-atom ensembles imposed as ideal candidates to quantum-simulate condensed-matter systems \cite{BlochNatPhys12}.
In this perspective a fundamental role is played by the interparticle potential, usually described in terms of a contact interaction, which can be tuned at will exploiting Feshbach resonances \cite{ChinRMP10}.
Two-body scattering is strongly modified by reducing the dimensionality of the system, leading to confinement-induced resonances (CIR) \cite{OlshaniiPRL98}.
Furthermore, the dipolar interaction, due to its long-range and anisotropic character, is a promising candidate to mimic more general Hamiltonians \cite{LahayeRPP09}.
The investigation of dipolar quantum gases has been boosted by the realization of Bose-Einstein condensates of magnetic dipoles (namely, Cr, Er, and Dy \cite{DipolarBEC}) and by the recent progresses with heteronuclear molecules (such as RbK and NaK \cite{Heteronuclear}).
A fundamental feature of the low-energy scattering between either magnetic or electric dipoles is the occurrence of dipolar-induced resonances (DIRs) when the dipole strength is varied~\cite{MarinescuPRL98}.

\begin{figure}
\begin{center}
\includegraphics[width=.95\textwidth]{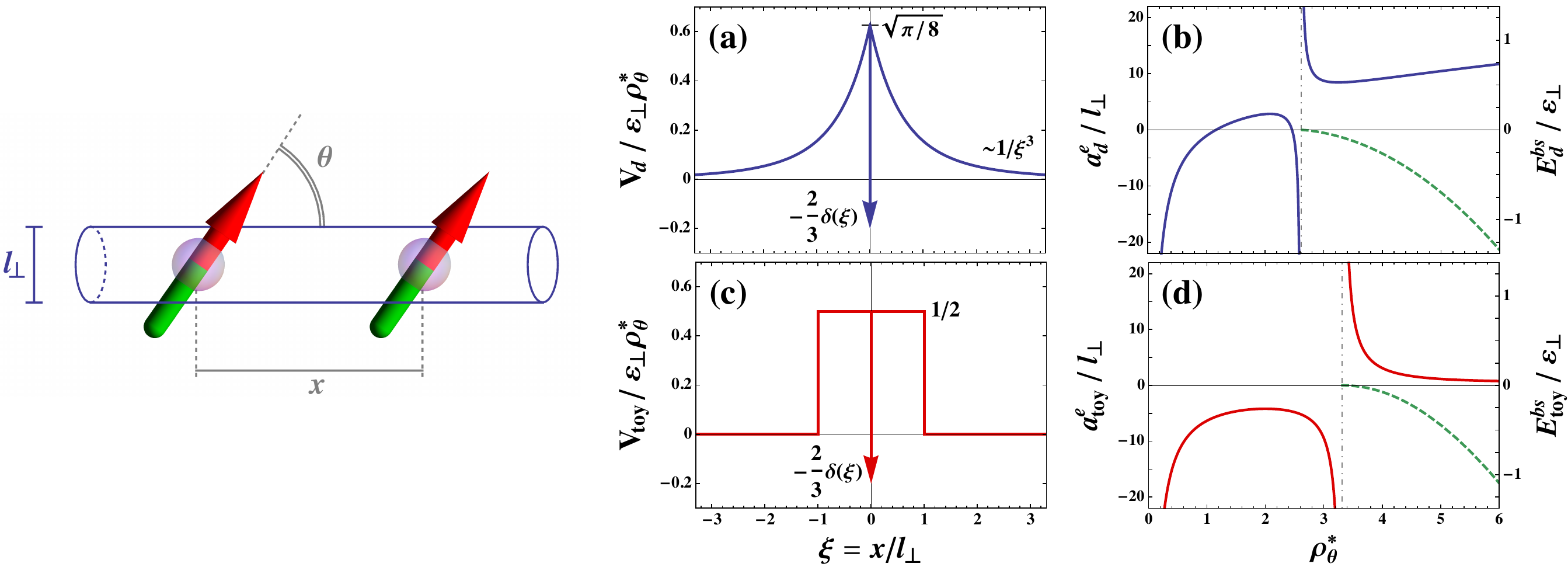}
\caption{(Color online)
Left: Representation of two polarized dipoles at relative distance $x$, harmonically trapped along a quasi-1D tube, of radial size $l_{\!\perp}=(\hbar/m\omega_{\!\perp})^{1/2}$.
The polarization direction and the tube axis form an angle $\theta$.
(a) Effective 1D dipolar interaction $V_{\rm d}$ [Eq.\eqref{DefVDI}] and (c) corresponding toy-model potential $V_{\rm toy}$ [Eq.\eqref{DefVtoy}].
(b) Numerical results for the even-channel scattering length $a_{\rm d}^e$ associated to $V_{\rm d}$ [solid (blue)], as obtained from Eq.~\eqref{Defa} for $x\!=\!100l_{\!\perp}$.
A single DIR occurs for $\rrho\simeq2.6$, in coincidence with the entrance of a dipolar bound state of energy $E_{\rm d}^{bs}$ [dashed (green)].
(d) Even-channel scattering length $a_{\rm toy}^e$ of the toy model.
The qualitative behavior is the same of $a_{\rm d}^e$, with a resonance occurring for $\rrho\simeq3.3$, coinciding with the entrance of a bound state of energy $E_{\rm toy}^{bs}$ [dashed (green)].
\label{FigSystem}}
\end{center}
\end{figure}

In this work we consider the two-body low-energy scattering of polarized dipoles tightly confined along a quasi-one-dimensional (quasi-1D) tube \cite{BartoloPRA13,GiannakeasPRL13}, represented in Fig.~\ref{FigSystem}(left).
As described in Sec.~\ref{SecTube}, a dipolar-induced resonance (DIR) emerges in this system.
It is practical to investigate the scattering problem by introducing a toy-model potential that catches the main features of the dipolar interaction (Sec.~\ref{SecToy}).
The toy-model scattering length can be calculated analytically and shows a resonant behavior analogous to that of the DIR.
In Sec.~\ref{SecToyGen}, the toy model is finally generalized to investigate the interplay of contact and non-zero-range potentials, giving an intuitive interpretation of the mechanism responsible for the resonance.

\section{Dipolar-induced resonance in a quasi-1D tube}\label{SecTube}

Generally, the dipolar interaction not only depends on the inter-particle distance, but also on the relative orientation of dipoles in space \cite{LahayeRPP09}.
It is considerably simplified in the case of polarized dipoles, a scheme realizable by aligning the particles with an external field.
In this case it takes the form $V_{\rm d}^{\rm 3D}\!=\!\hbar^2 r^* (1-3\cos^2\theta)/(mr^3)$,
where we have introduced the dipolar length $r^*\!=\!mD^2/\hbar^2$
(being $D$ the dipole strength and $m$ its mass).
The system we want to investigate, represented in Fig.~\ref{FigSystem}(left), consists of two polarized dipoles in a quasi-1D tube.
The particles are harmonically confined in the radial direction with trapping frequency $\omega_{\!\perp}$, which fixes the length $l_{\!\perp}\!=\!(\hbar/m\omega_{\!\perp})^{1/2}$.
Two classical dipoles restricted to 1D motion would simply repel (attract) each other for $\theta_c\!<\!\theta\!<\!\pi/2$ ($0\!<\!\theta\!<\!\theta_c$), where $\theta_c\!\simeq\!54.7^\circ$ is the ``magic angle'' at which the interaction vanishes.
In the quantum case, we need to account for the radial extension of the particles' wavefunctions.
By restricting the analysis to scattering energies $E\!\ll\!\varepsilon_{\!\perp}\!=\!\hbar\omega_{\!\perp}$, we assume the dipoles to lie in the transverse ground state.
Properly integrating out the transverse degrees of freedom \cite{EffectivePot}, one can consider the tube as an effective 1D system, in which $V_{\rm d}^{\rm 3D}$ is replaced by the effective relative potential
\begin{equation}\label{DefVDI}
V_{\rm d}(x)=\varepsilon_{\!\perp} \rho_\theta^*\,\left[ w\left(   \frac{x}{l_{\!\perp}} \right)-\frac{2}{3} \delta \left(\frac{x}{l_{\!\perp}} \right) \right],
\end{equation}
with  $\rrho\!=\!r^*(1-3\cos^2\theta)/l_{\!\perp}$ and
$w(\xi)\!=\!\sqrt{\pi/8}\, (1+\xi^2)\, \exp(\xi^2/2)\,
{\rm erfc}(|\xi|/\sqrt{2})-|\xi|/2$.
The potential $V_{\rm d}$ is plotted in Fig.~\ref{FigSystem}(a) for $\rrho\!>\!0$, corresponding to the regime of classical repulsion.
Distant particles perceive each other as classical dipoles, so that the potential has the expected $1/x^3$ behavior for $x\!\gg\!l_{\!\perp}$.
On the other hand, getting closer, the dipoles' quantum nature emerges: the interplay of transverse extension and interaction anisotropy results in a non-divergent repulsion plus an attractive contact term.
For $\rrho\!<\!0$ the potential is reversed, resulting mainly attractive, and for any small value of $r^*$ there exists at least one dipolar bound state.

The scattering properties of the 1D potential $V_{\rm d}(x)$ can be investigated by numerically solving the relative-motion zero-energy Schr\"odinger equation
\begin{equation}\label{1DSch}
\bigl[V_{\rm d}(x)-\hbar^2\partial_x^2/m\bigr]\psi_p(x)=0.
\end{equation}
The parity index $p\!=\!e,o$ distinguishes between even and odd solutions, corresponding, respectively, to bosonic and fermionic particles.
The $s$-wave scattering length for each channel is defined by
\begin{equation}\label{Defa}
a_{\rm d}^p\!=\!\lim_{x\to\infty}[x-\psi_p(x)/\psi_p'(x)].
\end{equation}
Due to the long-range character of $V_{\rm d}$, it is not possible to associate a well defined scattering length to it, since Eq.~\eqref{Defa} does not converge \cite{AstrakharchikPRA08}.
Anyhow, one can evaluate Eq.~\eqref{Defa} for a large, but finite value $x\!=\!x_{\rm max}$.
The even-channel scattering length of $V_{\rm d}$ for $x_{\rm max}\!=\!100l_{\!\perp}$ is presented in Fig.~\ref{FigSystem}(b).
A DIR occurs at $\rrho\!\simeq\!2.6$, due to the presence of the attractive $\delta$-term which takes over the repulsive wings, allowing for the existence of a dimer state of energy $E_{\rm d}^{bs}$ also in the regime of classical repulsion.
Remarkably, the resonance position is
unchanged for any $x_{\rm max}\!\gtrsim\!10l_{\!\perp}$.
Numerical estimations suggest that this is the only DIR for $\rrho\!>\!0$ in a quasi-1D tube.
The contact term is invisible to odd solutions, so that no resonances arise in this case.

\section{Toy model for the effective dipolar interaction}\label{SecToy}
To investigate the physical properties of $V_{\rm d}$, we propose a versatile toy model for which $a_{\rm toy}^p$ can be evaluated analytically, together with the energy $E_{\rm toy}^{bs}$ of the dimer state appearing at the resonance.
Hence, we replace the fast-decaying wings of the real potential with a finite-range step function:
\begin{equation}\label{DefVtoy}
V_{\rm toy}(x)=\varepsilon_{\!\perp}\rrho\, \left[ \frac{1}{2}\, \sigma\left(\frac{x}{l_{\!\perp}}\right) - \frac{2}{3}\, \delta\left(\frac{x}{l_{\!\perp}}\right) \right],
\quad{\rm with}\quad
\sigma\left(\frac{x}{l_{\!\perp}}\right)=\begin{cases}
1   &    |x|\le l_{\!\perp},  \\
0       &   |x|>l_{\!\perp}.
\end{cases}
\end{equation}
The step width $2l_{\!\perp}$ corresponds to the region in which $V_{\rm d}$ deviates from the classical $1/x^3$ behavior, while its height $\varepsilon_{\!\perp}/2$ has been chosen so that areas under the wings $w$ and the step $\sigma$ are the same.
The potential $V_{\rm toy}$ is plotted in Fig.~\ref{FigSystem}(c), together with the corresponding even-channel scattering length $a_{\rm toy}^e(\rrho)$ [Fig.~\ref{FigSystem}(d)].
The model is able to reproduce the DIR, with a resonance appearing at $\rrho\!\simeq\!3.3$ for even wave functions, which results close to the value $\rrho\!\simeq\!2.6$ found for the real potential, despite the simplicity of the model.
Furthermore, the analytic expression of $a_{\rm toy}^e$ [reported in Eq.~\eqref{DefaToyGen} for a more general case] confirms that only one DIR exists for $\rrho\!>\!0$.

\section{Interplay of dipolar and contact interaction}\label{SecToyGen}
In addition to the dipolar interaction, we now consider the presence of a contact potential, whose strength is fixed by the 3D scattering length $a_{\rm3D}$.
In a quasi-1D tube, this is taken into account by the effective contact potential
$V_{\rm c}(x)\!=\!2\varepsilon_{\!\perp}(a_{\rm3D}/l_{\!\perp})\delta(x/l_{\!\perp})$
\cite{PetrovPRL00}, valid as long as $a_{\rm3D}\!\ll\!l_{\!\perp}$.
Thus, in a real system, one can change independently the contact and long-range terms of the total interaction $V_{\rm d}+V_{\rm c}$ by tuning $a_{3\rm D}$ via a Feshbach resonance \cite{ChinRMP10} and changing $\rrho$ with the polarizing field \cite{LahayeRPP09}.
Correspondingly, we can generalize the toy-model potential
\begin{equation}\label{DefVtoyGen}
\widetilde{V}_{\rm toy}(x)=\varepsilon_{\!\perp}\, \left[ \beta\, \sigma\left(\frac{x}{l_{\!\perp}}\right) + \alpha\, \delta\left(\frac{x}{l_{\!\perp}}\right) \right],
\end{equation}
where the parameters $\alpha\!=\!2a_{\rm3D}/l_{\!\perp}-2\rrho/3$ and $\beta\!=\!\rrho/2$ set, respectively, the contact and non-zero-range interaction strengths.
By analytically solving the Schr\"odinger equation~\eqref{1DSch} for $\widetilde{V}_{\rm toy}$, one gets the scattering lengths:
\begin{equation}\label{DefaToyGen}
\widetilde{a}_{\rm toy}^{e}(\alpha,\beta)=1-\frac{1}{\kappa}
\frac{\alpha\sinh(\kappa)+2\kappa\cosh(\kappa)}{2\kappa\sinh(\kappa)+\alpha\cosh(\kappa)},
\quad
\widetilde{a}_{\rm toy}^{o}(\alpha,\beta)=1-\frac{\tanh(\kappa)}{\kappa},
\end{equation}
with $\kappa\!=\!\sqrt{\beta}$ for $\beta\!>\!0$ and $\kappa\!=\!i\sqrt{|\beta|}$ for $\beta\!<\!0$.

\begin{figure}
\begin{center}
\includegraphics[width=0.95\textwidth]{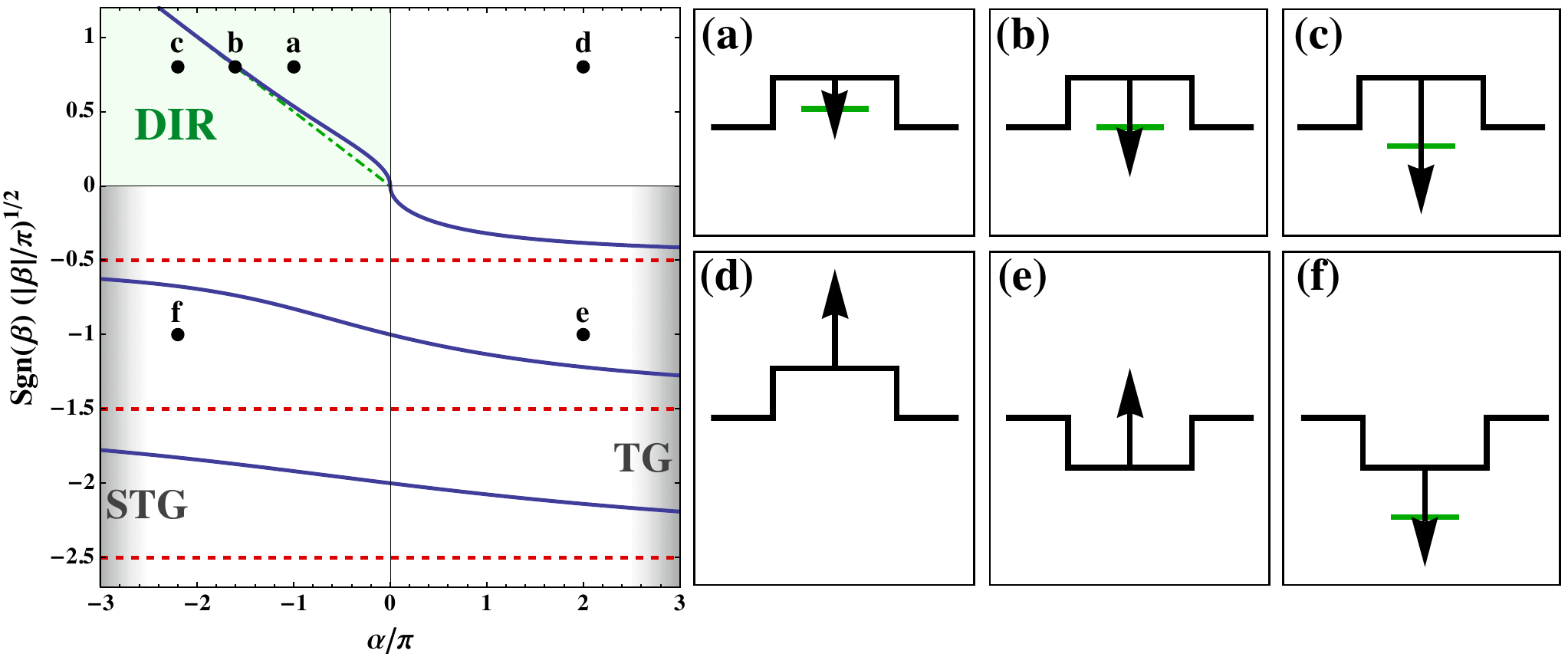}
\caption{(Color online)
Left: Solid blue (dashed red) lines correspond to solutions of $1/\widetilde{a}_{\rm toy}^e\!=\!0$ ($1/\widetilde{a}_{\rm toy}^o\!=\!0$) [cf. Eq.~\eqref{DefaToyGen}].
The green dot-dashed line marks the condition $E_{\!\alpha}\!=\!E_{\!\beta}$ (cf. text).
Gray-shadowed regions ($\beta\!<\!0$) indicate the direction in which the Tonks-Girardeau (TG) and super-Tonks-Girardeau (STG) limits are asymptotically reached.
The green-shadowed quadrant ($\alpha\!<\!0$, $\beta\!>\!0$) is the one in which the DIR mechanism occurs.
Right: Illustration of the generalized toy-model potential [Eq.~\eqref{DefVtoyGen}] at the points (a,b,c,d,e,f) marked on the resonances diagram at left.
Green horizontal lines represent the energy $E_\alpha$ of the $\delta$-sustained bound state, shifted upwards by the height of the repulsive energy barrier $E_\beta$.
\label{FigToyResonances}}
\end{center}
\end{figure}

In Fig.~\ref{FigToyResonances}(left) we show the position of the resonances of $\widetilde{a}_{\rm toy}^p$ varying $\alpha$ and $\beta$.
The contact term is invisible to odd solutions, since $\psi_o(0)\!=\!0$.
Hence, the corresponding resonances do not depend on $\alpha$.
They exist only for $\beta\!<\!0$ and are simply those of a square well of depth $|\beta|$.
Even solutions are, instead, strongly affected by the $\delta$-potential.
No resonances exist in the purely repulsive quadrant $\alpha,\beta\!>\!0$.
For $\alpha\!>\!0$ and $\beta\!<\!0$ [Fig.~\ref{FigToyResonances}(e)], when $\alpha\!\to\!\infty$ the system reaches the Tonks-Girardeau limit of impenetrable particles \cite{GirardeauJMP60}:
even wave functions acquire a zero at the origin to avoid a divergent contribution to the energy and, correspondingly, the even resonances tend asymptotically to the odd ones. 
A similar even-to-odd limit occurs in the region $\alpha,\beta\!<\!0$ [Fig.~\ref{FigToyResonances}(f)].
For $\alpha\!\to\!-\infty$ the $\delta$-potential sustains only a single, infinitely deep bound state, so that the other even wavefunctions must acquire a zero at the origin to keep finite their energy.
The dipoles become effectively impenetrable, reaching the super-Tonks-Girardeau regime \cite{SuperTonks}.
The DIR occurs if $\alpha\!<\!0$ and $\beta\!>\!0$.
It is a (single-channel) shape resonance which results from a competition between the attractive delta term and the repulsive step potential, and it can be understood intuitively as follows.
In the absence of the step potential ($\beta=0$), the $\delta$~term would support a bound state with energy $E_\alpha\!=\!-\varepsilon_{\!\perp}\alpha^2/4\!<\!0$. 
If we now add to the Hamiltonian a step potential whose height $E_\beta\!=\!\beta\varepsilon_{\!\perp}\!>\!0$ is smaller than $\sim\!|E_\alpha|$, the discrete level survives and its energy is shifted upwards by $\sim\!E_\beta$.
On the other hand, if $E_\beta\!\gtrsim\!|E_\alpha|$, the discrete level supported by the delta dissolves into the continuum and disappears.
The resonance occurs at the threshold between these two regimes, i.e. for $E_\beta\!\sim\!|E_\alpha|$.
Our results [Fig.~\ref{FigToyResonances}(left)] show that this condition is asymptotically exact (if $|\alpha|$ and $\beta$ are both large, the resonance occurs for $E_{\beta}\!=\!|E_{\alpha}|$).

\section{Conclusions and Perspectives}\label{SecConc}
We presented a short study on the dipolar interaction in a quasi-1D tube, discussing the emergence of a DIR and the appearance of a dimer state in a regime where classical dipoles would simply repel each other.
This feature can be reproduced using a simple toy-model potential for which the low-energy scattering properties are analytically determined.
A generalized version of the toy model allows to investigate the interplay of contact and step potential, giving an intuitive description of the DIR as a shape resonance.
These results can be extended for the actual quasi-1D dipolar plus contact potential $V_{\rm d}+V_{\rm c}$, which will be likely to present analogous features.
Furthermore, interesting effects may arise for $a_{\rm3D}\!\gtrsim\!l_{\!\perp}$, due to the coaction of DIR and CIR.
These investigations on the two-body scattering are the building block for the analysis of many-body dipolar systems in optical lattices \cite{BartoloPRA13} and are useful in determining the stability conditions for systems of attracting dipoles.
The effects of the DIR may be probed experimentally by studying the energy levels with spectroscopic techniques or by looking at the two-body losses as a function of the dipolar strength.

\smallskip
This work has been supported by ERC through the QGBE grant and by Provincia Autonoma di Trento.


\end{document}